\documentclass[proceedings]{stacs}
\stacsheading{2009}{457--468}{Freiburg}
\firstpageno{457}

\usepackage{psfrag}
\usepackage{graphicx}
\usepackage{mathrsfs}
\usepackage{amsfonts}
\usepackage{amssymb}

\newcommand{\PERM}{\textsc{perm}}
\newcommand{\SEL}{\textsc{bfprt}}

\newcommand{\junk}[1]{}

\newcommand{\reduced}{\mathscr{K}}
\newcommand{\act}[1]{\mathcal{A}_{#1}}
\newcommand{\inact}[1]{\mathcal{I}_{#1}}

\newcommand{\pivot}{p}

\newcommand{\A}{\ensuremath{\mathtt{a}}}
\newcommand{\B}{\ensuremath{\mathtt{b}}}
\newcommand{\M}{\ensuremath{\mathtt{\$}}}

\newcommand{\presel}[1]{\sigma_{#1}}
\newcommand{\prerank}[1]{z_{#1}}

\newcommand{\wo}[1]{w_{#1}}
\newcommand{\suf}[1]{s_{#1}}
\newcommand{\pro}[1]{p_{#1}}

\newcommand{\reach}[1]{r_{#1}}
\newcommand{\preach}[1]{{pr}_{#1}}
\newcommand{\tailpr}[1]{\mathit{tail}\left({\preach{#1}}\right)}
\newcommand{\multpreach}[1]{\mathit{mult}\left({\preach{#1}}\right)}

\newcommand{\collis}[1]{C}
\newcommand{\pseudocollis}[1]{PC\left({#1}\right)}

\newcommand{\fblock}{q}
\newcommand{\period}{u}
\newcommand{\compar}{Comp}
\newcommand{\prefi}[1]{p\suf{#1}}
\newcommand{\firstint}[1]{\alpha_{#1}}
\newcommand{\secondint}[1]{\beta_{#1}}
\newcommand{\ordrelation}{\vartriangleleft}
\newcommand{\pairs}{Pairs}
\newcommand{\pairscomp}{PComp}
\newcommand{\wk}{{\tt work}}

\newcommand{\set}[1]{\left\{#1\right\}}
\newcommand{\seq}[1]{\left\langle#1\right\rangle}

\newcommand{\card}[1]{\left\vert#1\right\vert}
\newcommand{\lcp}[1]{\mathit{lcp}\left({#1}\right)}

\newcommand{\Oh}[1]{O \left ( #1 \right )}
\newcommand{\Omegah}[1]{\Omega \left ( #1 \right )}
\newcommand{\Thetah}[1]{\Theta \left ( #1 \right )}

\newcommand{\omegah}[1]{\omega \left ( #1 \right )}

\newcommand{\Min}[1]{\mbox{min}\left\{#1\right\}}

\begin{document}
\title[Optimal Cache-Aware Suffix Selection]{Optimal Cache-Aware Suffix Selection}
\author[uniroma]{G. Franceschini}{Gianni Franceschini}
\address[uniroma]{Dipartimento di Informatica, Universit\`a di Roma ``La
Sapienza'', and PINT, Primorska University}
\email{francesc@di.uniroma1.it}
\author[unipi]{R. Grossi}{Roberto Grossi}
\address[unipi]{Dipartimento di Informatica, Universit\`a di Pisa}
\email{grossi@di.unipi.it}
\author[google]{S. Muthukrishnan}{S. Muthukrishnan}
\address[google]{Google Inc., NY}
\email{muthu@google.com}

\thanks{The first two authors have been partially supported by the MIUR project MAINSTREAM}







\def\spacingafterparagraph{\paragraph{}\hskip-.5em\null\indent }
\setlength{\parskip}{-0.3ex}

\newtheorem{property}{Property}

\def\boxit#1{\vbox{\hrule\hbox{\vrule\kern3pt
  \vbox{\kern3pt#1\kern3pt}\kern3pt\vrule}\hrule}} 
\def\Box{\rule{2mm}{3mm}} 

\newenvironment{proofof}[1]{\trivlist\item[]{\em Proof of #1\/}:}%
{\unskip\nobreak\hskip 2em plus 1fil\nobreak$\Box$
\parfillskip=0pt%
\endtrivlist}                                                                              



\newcommand{\bsubsection}[1]{%
\vspace*{-.3ex}
\paragraph*[#1]{\bf #1}\ignorespaces
\vspace*{-.3ex}}

\newcommand{\iparagraph}[1]{%
\vspace*{-.3ex}
\paragraph*[#1]{\it #1}\ignorespaces
\vspace*{-.3ex}}

\newenvironment{enumroman}
{\renewcommand{\theenumi}{(\roman{enumi})}
\renewcommand{\labelenumi}{\theenumi}
\begin{enumerate}}
{\end{enumerate}}


\newenvironment{enumalpha}
{\renewcommand{\theenumi}{(\alph{enumi})}
\renewcommand{\labelenumi}{\theenumi}
\begin{enumerate}}
{\end{enumerate}}


\begin{abstract}
  Given string $S[1..N]$ and integer $k$, the {\em suffix selection}
  problem is to determine the $k$th lexicographically smallest amongst
  the suffixes $S[i\ldots N]$, $1 \leq i \leq N$.  We study the suffix
  selection problem in the cache-aware model that captures two-level
  memory inherent in computing systems, for a \emph{cache} of limited
  size $M$ and block size $B$. The complexity of interest is the
  number of block transfers.  We present an optimal suffix selection
  algorithm in the cache-aware model, requiring $\Thetah{N/B}$ block
  transfers, for any string $S$ over an unbounded alphabet (where
  characters can only be compared), under the common tall-cache
  assumption (i.e. $M=\Omegah{B^{1+\epsilon}}$, where $\epsilon<1$).
  Our algorithm beats the bottleneck bound for permuting an input
  array to the desired output array, which holds for nearly any
  nontrivial problem in hierarchical memory models.
\end{abstract}
\def\today{}
\maketitle


\section{Introduction}
\label{sec:introduction}

\bsubsection{Background: Selection vs Sorting}

A collection of $N$ {\em numbers} can be sorted using $\Thetah{N\log N}$
comparisons. On the other hand, the famous five-author
result~\cite{BFPRT73} from early 70's shows that the problem of {\em
  selection} --- choosing the $k$th smallest number --- can be solved using
$\Oh{N}$ comparisons in
the worst case. Thus, selection is provably simpler than sorting in
the comparison model. 

Consider a sorting vs selection question for strings.  Say
$S=S[1\cdots N]$ is a string. The {\em suffix sorting} problem is to
sort the suffixes $S[i\cdots N]$, $i=1,\ldots,N$, in the {\em
  lexicographic} order. In the comparison model, we count the number
of character comparisons. 
Suffix
sorting can be performed with $\Oh{N\log N}$ comparisons using a
combination of character sorting and classical data structure of
suffix arrays or trees~\cite{W73,McC76,Far97}.  There is a
lower bound of $\Omegah{N\log N}$ since sorting suffixes ends up
sorting the characters.  For 
the related {\em suffix selection} problem where the goal is to
output the $k$th lexicographically smallest 
suffix of $S$, the result in~\cite{FM07a} recently gave an optimal $\Oh{N}$
comparison-based algorithm, thereby showing that
suffix selection is provably simpler than suffix sorting.

\bsubsection{The Model}

Time-tested architectural approaches to computing
systems provide two (or more) levels of memory: the highest one with a limited
amount of fast memory; the lowest one with slow but large memory. The
CPU
can only access input stored on the fastest level. Thus, there is a continuous
exchange of
data between the levels. For cost and performance reasons, data is exchanged in 
fixed-size blocks of contiguous locations. These transfers may be 
triggered automatically like in internal CPU caches, or explicitly,
like in the case of disks; in either case, more than the number of computing
operations
executed, the number of block transfers required is the actual bottleneck.

Formally, we consider the model that has two memory levels. 
The \emph{cache} level contains $M$ locations
divided into blocks (or cache lines) of $B$ contiguous locations, and the
\emph{main memory} level can be arbitrarily large and is also divided into
blocks. The processing unit can
address the locations of the main memory but it can process only the
data residing in cache. The algorithms that know and exploit the two
parameter $M$ and $B$, and optimize the number of block transfers are 
\emph{cache-aware}. This model includes the classical External Memory model
\cite{AV88}
as well as the well-known \emph{Ideal-Cache model}
\cite{FLPR99}. 

\bsubsection{Motivation}

 
Suffix selection as a problem is useful in analyzing the order statistics of
suffixes in a string such as the extremes, medians and outliers, with
potential applications in bioinformatics and information retrieval.  A
quick method for finding say the suffixes of rank $i (n/10)$ for each
integer $i$, $0\leq i \leq 10$, may be used to partition the space of
suffixes for understanding the string better, load balancing and
parallelization.  But in these applications, such as in bioinformatics, the
strings are truly massive and unlikely to fit in the fastest levels of memory.
Therefore it is 
natural to analyze them in a hierarchical memory model.

Our primary motivation however is really theoretical. Since the inception of
the first block-based hierarchical memory model 
(\cite{AV88},\cite{vit}), 
it has been difficult to obtain  ``golden
standard'' algorithms  i.e., those using just $\Oh{N/B}$ block transfers.
Even the simplest permutation problem (\PERM\, henceforth) where the
output is a specified permutation of the input array, does not have
such an algorithm. In the standard RAM model, \PERM\ can be solved in
$\Oh{N}$ time. In both the Ideal-Cache and External Memory models, the
complexity of this problem is denoted \PERM$(N) =
\Thetah{\Min{N,(N/B\log_{M/B} N/B}}$.
Nearly any nontrivial problem one can imagine from list ranking to
graph problems such as Euler tours, DFS, connected components etc.,
sorting and  geometric problems have the lower bound of
\PERM$(N)$, even if they take $\Oh{N}$ time in the RAM model, and
therefore do not meet the ``golden standard''. Thus the lower bound
for \PERM\ is a terrible bottleneck for block-based hierarchical memory models.

The outstanding question is, much as in the comparison model, 
is suffix selection provably simpler than 
suffix sorting in the block-based hierarchical memory models?
Suffix sorting takes $\Thetah{(N/B)\log_{M/B} (N/B)}$
block transfers~\cite{FM}. Proving any problem to be simpler than 
suffix sorting therefore requires one to essentially overcome the \PERM\
bottleneck. 

\bsubsection{Our Contribution}

We present a suffix selection algorithm with optimal cache complexity.
Our algorithm requires
$\Thetah{N/B}$ block transfers, for any string $S$ over an
unbounded alphabet (where characters can only be compared) and under the common
tall-cache assumption, that is $M=\Omegah{B^{1+\epsilon}}$ with
$\epsilon<1$. 
Hence, we meet the ``golden standard''; we beat the \PERM\ bottleneck and
consequently, prove that suffix selection is easier than suffix sorting in
block-based hierarchical memory models.



\bsubsection{Overview}

Our high level strategy for achieving an optimal cache-aware suffix selection
algorithm consists of two main objectives.

In the first objective, we want to efficiently reduce the number of
candidate suffixes from $N$ to $\Oh{N/B}$, where we maintain the
invariant that the wanted $k$th smallest suffix is surely one of the
candidate suffixes.


In the second objective, we want to achieve a cache optimal solution
for the \emph{sparse suffix selection problem}, where we are given a
subset of $\Oh{N/B}$ suffixes including also the wanted $k$th
suffix. To achieve this objective we first find a simpler approach to
suffix selection for the standard comparison model. (The only known
linear time suffix selection algorithm for the comparison model
\cite{FM07a} hinges on well-known algorithmic and data structural
primitives whose solutions are inherently cache inefficient.)  Then,
we modify the simpler comparison-based suffix selection algorithm to
exploit, in a cache-efficient way, the hypothesis that $\Oh{N/B}$
(known) suffixes are the only plausible candidates.


\iparagraph{Map of the paper.}
We  will start by
describing the new simple
comparison-based suffix
selection algorithm in Section~\ref{sec:simple}. This section is meant to be intuitive. We will use 
it to derive a cache-aware algorithm for the sparse suffix
selection problem in Section~\ref{sec:sparse:selection}.  We will
present our optimal cache-aware algorithm for the general suffix
selection problem in Section~\ref{sec:overview-external-memory}.

\section{A Simple(r) Linear-Time Suffix Selection Algorithm}
\label{sec:simple}

We now describe a simple algorithm for selecting the
$k$th lexicographically smallest suffix of~$S$ in main memory.
We give some intuitions on the central notion of \emph{\wk{}}, and some definitions
and notations used in the algorithm. Next, we show how to perform 
main iteration, called \emph{phase transition}.  Finally, we present
the invariants that are maintained in each phase transition, and
discuss the correctness and the complexity of our algorithm.

\bsubsection{Notation and intuition}

Consider the regular linear-time selection algorithm~\cite{BFPRT73},
hereafter called \SEL.  Our algorithm for a string $S = S[1\ldots N]$
uses \SEL\ as a black box.\footnote{In the following, we will assume
  that the last symbol in $S$ is an endmarker $S[N] = \M$, smaller
  than any other symbol in $S$.}  Each run of \SEL\ permits to
discover a longer and longer prefix of the (unknown) $k$th
lexicographically smallest suffix of $S$. We need to carefully
orchestrate the several runs of \SEL\ to obtain a total cost of $O(N)$
time. We use $S = \B\B\B\A\B\B\B\B\B\A\A\M$, where $n=12$, as an
illustrative example, and show how to find the median suffix (hence,
$k=n/2=6$).

\iparagraph{Phases and phase transitions.}
We organize our computation so that it goes through \emph{phases},
numbered $t=0,1,2,\ldots$ and so on. In phase $t$, we know that a
certain string, denoted $\presel{t}$, is a prefix of the (unknown)
$k$th lexicographically smallest suffix of $S$. Phase $t=0$ is the
initial one: we just have the input string $S$ and no knowledge, i.e.,
$\presel{0}$ is the empty string.  For $t \geq 1$, a main iteration of
our algorithm goes from phase~$t-1$ to phase $t$ and is termed
\emph{phase transition} $(t-1 \rightarrow t)$: it is built around the
$t$th run of \SEL\ on a suitable subset of the suffixes of $S$. Note
that $t \leq N$, since we ensure that the condition $|\presel{t-1}| <
|\presel{t}|$ holds, namely, each phase transition extends the known
prefix by at least one symbol.

\iparagraph{Phase transition $(0 \rightarrow 1)$.}
We start out with phase~0, where we run \SEL\ on the individual
symbols of~$S$, and find the symbol $\alpha$ of rank $k$ in $S$ (seen
as a multiset). Hence we know that $\presel{1} = \alpha$, and this
fact has some implications on the set of suffixes of $S$.  Let
$\suf{i}$ denote the $i$th suffix $S[i\ldots N]$ of $S$, for $1 \le i
\le N$, and $\wo{i}$ be a special prefix of $\suf{i}$ called 
\emph{\wk{}}. We anticipate that the \wk{}s play a fundamental role in
attaining $O(N)$ time. To complete the phase transition, we set
$\wo{i} = S[i]$ for $1 \leq i \le N$, and we call \emph{degenerate}
the \wk{}s $\wo{i}$ such that $\wo{i} \ne \alpha$. (Note that degenerate
\wk{}s are only created in this phase transition.)  We then partition
the suffixes of $S$ into two disjoint sets:
\begin{itemize}
\item The set of {\em active suffixes}, denoted by $\act{1}$---they
  are those suffixes $\suf{i}$ such that $\wo{i} = \presel{1} = \alpha$.
\item The set of {\em inactive suffixes}, denoted by $\inact{1}$ and
  containing the rest of the suffixes---none of them is surely the
  $k$th lexicographically smallest suffix in $S$.
\end{itemize}

In our example ($k=6$), we have $\presel{1} = \alpha = \B$ and, for $i
= 1,2,3,5,6,7,8,9$, $\wo{i}=\B$ and $\suf{i} \in \act{1}$. Also, we
have $\suf{j} \in \inact{1}$ for $j = 4, 10, 11, 12$, where $\wo{4}
= \wo{10} = \wo{11} = \A$ and $\wo{12} = \M$ are degenerate
\wk{}s.

A comment is in order at this point. We can compare any two \wk{}s in
constant time, where the outcome of the comparison is ternary
$[<,=,>]$. While this observation is straightforward for this phase
transition, we will be able to extend it to longer \wk{}s in the
subsequent transitions.  Let us discuss the transition from phase $1$
to phase $2$ to introduce the reader to the main point of the
algorithm.

\iparagraph{Phase transition $(1 \rightarrow 2)$.}
If $\card{\act{1}}=1$, we are done since there is only one active
suffix and this should be the $k$th smallest suffix in $S$.
Otherwise, we exploit the total order on the current \wk{}s.
Letting $\prerank{1}$ be the number of \wk{}s smaller than the current
prefix $\presel{1}$, our goal becomes how to find the
$(k-\prerank{1})$th smallest suffix in $\act{1}$.  In particular, we
want a longer prefix $\presel{2}$ and the new set $\act{2} \subseteq
\act{1}$. 

To this end, we need to extend some of the \wk{}s of the active
suffixes in $\act{1}$.  Consider a suffix $\suf{i} \in \act{1}$. In
order to extend its \wk{} $\wo{i}$, we introduce its
\emph{prospective \wk{}}.  Recall that $\wo{i} = \presel{1} = \alpha
= S[i]$. If $\wo{i+1} = S[i+1] \neq \alpha$ (hence, $\suf{i+1}$ is
inactive in our terminology), the prospective \wk{} for $\suf{i}$ is
the concatenation $\wo{i} \wo{i+1}$, where $\suf{i+1} \in
\inact{1}$. Otherwise, since $\wo{i} = \wo{i+1}$ (and so
$\suf{i+1} \in \act{1}$), we consider $i+2, i+3$, and so on, until we
find the first $i+r$ such that $\wo{i} \neq \wo{i+r}$ (and so
$\suf{i+r} \in \inact{1}$). In the latter case, the prospective \wk{}
for $\suf{i}$ is the concatenation $\wo{i} \wo{i+1} \cdots
\wo{i+r}$, where $\wo{i} = \wo{i+1} = \cdots = \wo{i+r-1} =
\presel{1} = \alpha$ and their corresponding suffixes are active, while
$\wo{i+r} \neq \presel{1}$ is different and corresponds to an
inactive suffix.

In any case, each prospective \wk{} is a sequence of \wk{}s of the form
$\alpha^r \beta = \presel{1}^r \beta$, where $r \ge 1$ and $\beta \ne
\alpha$. The reader should convince herself that any two prospective
\wk{}s can be compared in $\Oh{1}$ time. We exploit this fact by
running \SEL\ on the set $\act{1}$ of active suffixes and, whenever
\SEL\ requires to compare any two $\suf{i}, \suf{j} \in \act{1}$, we
compare their prospective \wk{}s. Running time is therefore
$O(\card{\act{1}})$ if we note that prospective \wk{}s can be easily
identified by a scan of $\act{1}$: if $\wo{i} \wo{i+1} \cdots
\wo{i+r}$ is the prospective \wk{} for $\suf{i}$, then $\wo{i+1}
\cdots \wo{i+r}$ is the prospective \wk{} for $\suf{i+1}$, and so
on. In other words, a consecutive run of prospective \wk{}s forms a
\emph{collision}, which is informally a maximal concatenated sequence
of \wk{}s equal to $\presel{1}$ terminated by a \wk{} different from
$\presel{1}$ (this notion will be described formally in
Section~\ref{subsec:phase-transition-main}).

After \SEL\ completes its execution, we know the prospective \wk{} that
is a prefix of the (unknown) $k$th suffix in $S$. That prospective
\wk{} becomes $\presel{2}$ and $\act{2}$ is made up of the the suffixes
in $\act{1}$ such that their prospective \wk{} equals $\presel{2}$ (and
we also set $\prerank{2}$).

In our example, $\prerank{1} = 3$, and so we look for the third
smaller suffix in $\act{1}$. We have the following prospective \wk{}s:
one collision is made up of $\pro{1} = \B\B\B\A$, $\pro{2} = \B\B\A$,
and $\pro{3} = \B\A$; another collision is made up of $\pro{5} =
\B\B\B\B\B\A$, $\pro{6} = \B\B\B\B\A$, $\pro{7} = \B\B\B\A$, $\pro{8}
= \B\B\A$, and $\pro{9} = \B\A$. Algorithm \SEL\ discovers that
$\B\B\A$ is the third prospective \wk{} among them, and so $\presel{2}
= \B\B\A$ and $\act{2} = \{ \suf{2}, \suf{8} \}$ (and $\prerank{2} =
5$).

\iparagraph{How to maintain the \wk{}s.}
Now comes the key point in our algorithm. For each suffix $\suf{i} \in
\act{2}$, we update its \wk{} to be $\wo{i} = \presel{2}$ (whereas it
was $\wo{i} = \presel{1}$ in the previous phase transition, so it is
now longer). For each suffix $\suf{i} \in \act{1} - \act{2}$, instead,
we leave its \wk{} $\wo{i}$ \emph{unchanged}. Note this is the key
point: although $\suf{i}$ can share a longer prefix with $\presel{2}$,
the algorithm \SEL\ has indirectly established that $\suf{i}$ cannot
have $\presel{2}$ as a prefix, and we just need to record a Boolean
value for $\wo{i}$, indicating if $\wo{i}$ is either
lexicographically smaller or larger than $\presel{2}$. We can stick to
$\wo{i}$ unchanged, and discard its prospective \wk{}, since
$\suf{i}$ becomes inactive and is added to
$\inact{2}$.
In our example, $\wo{2} = \wo{8} =
\B\B\A$, while the other \wk{}s are unchanged (i.e, $\wo{3} = \B$
while $\pro{3} = \B\A$, $\wo{5} = \B$ while $\pro{5}=\B\B\B\B\B\A$,
and so on).

In this way, we can maintain a \emph{total order on the \wk{}s}. If two
\wk{}s are of equal length, we declare that they are equal according to
the symbol comparisons that we have performed so far, unless they are
degenerate---in the latter case they can be easily compared as single
symbols. If two \wk{}s are of different length, say $\card{\wo{i}} <
\card{\wo{j}}$, then $\suf{i}$ has been discarded by \SEL\ in favor
of $\suf{j}$ in a certain phase, so we surely know which one is
smaller or larger. In other words, when we declare two \wk{}s to be
equal, we have not yet gathered enough symbol comparisons to
distinguish among their corresponding suffixes. Otherwise, we have
been able to implicitly distinguish among their corresponding
suffixes. In our example, $\wo{3} < \wo{2}$ because they are of
different length and \SEL\ has established this disequality, while we
declare that $\wo{3} = \wo{5}$ since they have the same
length. Recall that the total order on the \wk{}s is needed for
comparing any two prospective \wk{}s in $\Oh{1}$ time as we proceed in
the phase transitions.  The \wk{}s exhibit some other strong properties
that we point out in the invariants described in Section~\ref{subsec:inv}.

\iparagraph{Time complexity.}
From the above discussion, we spend $\Oh{\card{\act{1})}}$ time for
phase transition $(1 \rightarrow 2)$. We present a charging scheme to pay for that. \wk{}s
come again into play for an amortized cost analysis. Suppose that, in
phase~$0$, we initially assign each suffix $\suf{i}$ two kinds of
credits to be charged as follows: $O(1)$ credits of the first kind
when $\suf{i}$ becomes inactive, and further $O(1)$ credits of the
second kind when $\suf{i}$ is already inactive but its \wk{} $\wo{i}$
becomes the terminator of the prospective \wk{} of an active
suffix. Note that $\wo{i}$ is incapsulated by the prospective \wk{}
of that suffix (which survives and becomes part of $\act{2}$).

Now, when executing \SEL\ on $\act{1}$ as mentioned above, we have
that at most one prospective \wk{} \emph{survives} in each collision
and the corresponding suffix becomes part of $\act{2}$. We therefore
charge the cost $O(\card{\act{1}})$ as follows.  We take
$\Thetah{\card{\act{1}}-\card{\act{2}}}$ credits of the first kind
from the $\card{\act{1}}-\card{\act{2}} \geq 0$ active suffixes that
become inactive at the end of the phase transition. We also take
$\Thetah{\card{\act{2}}}$ credits from the $\card{\act{2}}$ inactive
suffixes whose \wk{} terminates the prospective \wk{} of the
survivors. In our example, the
$\Thetah{\card{\act{1}}-\card{\act{2}}}$ credits are taken from
$\suf{1}, \suf{3}, \suf{5}, \suf{6}, \suf{7}$, and $\suf{9}$, while
$\Thetah{\card{\act{2}}}$ credits are taken from $\suf{4}$ and
$\suf{10}$.

At this point, it should be clear that, in our example, the next phase
transition $(2\rightarrow 3)$ looks for the $(k-\prerank{2})$th
smaller suffix in $\act{2}$ by executing \SEL\ in $O(\card{\act{2}})$
time on the prospective \wk{}s built with the runs of consecutive 
occurrences of the \wk{} $\presel{2} = \B\B\A$ into $S$. We thus
identify $\B\B\A\A\M$ (with $\presel{3}=\B\B\A\A$) as the median
suffix in $S$.

\bsubsection{\boldmath Phase transition $(t-1 \rightarrow t)$ for $t \ge 1$}
\label{subsec:phase-transition-main}

We are now ready to describe the generic phase transition $(t-1
\rightarrow t)$ more formally in terms of the active suffixes in
$\act{t-1}$ and the inactive ones in $\inact{t-1}$, where $t \ge 1$.

The input for the phase transition is the following: (a)~the current
prefix $\presel{t-1}$ of the (unknown) $k$th lexicographically
smallest suffix in $S$; (b)~the set $\act{t-1}$ of currently active
suffixes; (c)~the number $\prerank{t-1}$ of suffixes in $\inact{t-1}$
whose \wk{} is smaller than that of the suffixes in $\act{t-1}$ (hence,
we have to find the $(k-\prerank{t-1})$th smallest suffix in
$\act{t-1}$); and (d)~a Boolean vector whose $i$th element is false
(resp., true) iff, for suffix $\suf{i} \in \inact{t-1}$, the algorithm
\SEL\ has determined that its \wk{} $\wo{i}$ is smaller (resp.,
larger) than $\presel{t-1}$. The output of the phase transition are
data (a)--(d) above, updated for phase $t$.

We now define collisions and prospective \wk{}s in a formal way.  We
say that two suffixes $\suf{i},\suf{j}\in\act{t}$ \emph{collide} if
their \wk{}s $\wo{i}$ and $\wo{j}$ are adjacent as substrings in
$S$, namely, $\card{i-j} = \card{\wo{i}} = \card{\wo{j}}$.  A {\em
  collision} $\collis{l}$ is the maximal subsequence $\wo{l_1}
\wo{l_2}\cdots\wo{l_r}$, such that
$\wo{l_1}=\wo{l_2}=\cdots=\wo{l_r}=\presel{t}$, where the active
suffixes $\suf{l_f}$ and $\suf{l_{f+1}}$ collide for any $1\le f<r$.
For our algorithm, a \emph{collision} can also be a degenerate
sequence of just one active suffix~$\suf{i}$ (since its \wk{} does not
collide with that of any other active suffix).

The {\em prospective \wk{}} of a suffix $\suf{i}\in\act{t-1}$, denoted
by $\pro{i}$, is defined as follows. Consider the collision
$\collis{l}$ to which $\suf{i}$ belongs. Suppose that $\suf{i}$ is the
$h$th active suffix (from the left) in $\collis{l}$, that is,
$\collis{l}=\wo{l_1}\wo{l_2}\cdots\wo{l_{h-1}}\wo{i}\wo{l_{h+1}}
\cdots\wo{l_{r-1}}\wo { l_r }$. Consider the suffix
$\suf{u}\in\inact{t-1}$ adjacent to $\wo{l_r}$ (because of the
definition of collision, $\suf{u}$ must be an inactive suffix
following $\wo{l_r}$). We define the prospective \wk{} of $\suf{i}$,
to be the string $\pro{i} =
\wo{i}\wo{l_{h+1}}\cdots\wo{l_{r-1}}\wo{l_r}\wo{u}$. Note
that $\wo{i} = \wo{l_{h+1}} = \cdots = \wo{l_{r-1}}= \wo{l_r}
= \presel{t-1}$ since their corresponding suffixes are all active,
while $\wo{u}$ is shorter. In other words, $\pro{i} =
\presel{t-1}^{r-h} \wo{u}$, with $\card{\wo{u}} <
\card{\presel{t-1}}$.

\begin{lemma}\label{lem:prospext:comp}
  For any two suffixes $\suf{i},\suf{j}\in\act{t}$, we can compare
  their prospective \wk{}s $\pro{i}$ and $\pro{j}$ in $\Oh{1}$ time. 
\end{lemma}

We now give the steps for the phase transition. Note that we can
maintain $\act{t-1}$ in monotone order of suffix position (i.e., $i <
j$ implies that $\suf{i}$ comes first than $\suf{j}$ in $\act{t-1}$).

\begin{enumerate}
\item\label{step:prosp}\label{step:let:T} Scan the active set
  $\act{t-1}$ and identify its collisions and the set $\mathscr{T}$
  containing all the suffixes $\suf{u}\in\inact{t-1}$ such that
  $\wo{u}$ immediately follows a collision. For any suffix $\suf{i}$
  in $\act{t-1}$, determine its prospective \wk{} $\pro{i}$ using the
  collisions and $\mathscr{T}$.
  
\item\label{step:select} Apply algorithm \SEL\ to the set
  $\set{\pro{i}}_{\suf{i}\in\act{t-1}}$ using the constant-time
  comparison as stated in Lemma~\ref{lem:prospext:comp}. In this way,
  find the $(k-\prerank{t-1})$th lexicographically smallest
  prospective \wk{} $p$, and the corresponding set
  $\act{t}=\set{\suf{i} \in \act{t-1} \;\vert\; \pro{i}=p}$ of active
  suffixes whose prospective \wk{}s match $p$.

\begin{enumalpha}
\item If $\card{\act{t}}=1$, stop the computation and return the
  singleton $\suf{i}\in\act{t}$ as the $k$th smallest suffix in $S$.
\item If $\card{\act{t}}>1$, set $\presel{t}=p$ (and update
  $\prerank{t}$ accordingly).
\end{enumalpha}

\item\label{step:set:act} For each $\suf{i}\in\act{t}$: Let
  $p=\wo{i}\wo{l_{h+1}} \cdots\wo{l_r}\wo{u}$ be its
  prospective \wk{}, where $\suf{u} \in \mathscr{T}$.  Set its new 
    \wk{} to be $\wo{i} = p = \presel{t}$.

\item\label{step:set:U:M} For each $\suf{j}\in\act{t-1}-\act{t}$,
  leave its \wk{} $\wo{j}$ \emph{unchanged} and, as a byproduct
  of running \SEL\ in step~\ref{step:select}, update position $j$  of the
Boolean
  vector (d) given in input, so as to record the fact that
  $\wo{j}$ is lexicographically smaller or larger than $\presel{t}$.
\end{enumerate}

\begin{lemma}\label{lem:transphase}
  Executing phase transition $(t-1 \rightarrow t)$ with $t \ge 1$,
  requires $\Oh{\card{\act{t-1}}}$ time in the worst case.
\end{lemma}

\bsubsection{\boldmath Invariants for phase $t$}
\label{subsec:inv}

Before proving the correctness and the complexity of our algorithm, we
need to establish some invariants that are maintained through the
phase transitions. We say that $\wo{i}$ is \emph{maximal} if there
does not exist another suffix $\suf{j}$ such that $\wo{j}$ contains
$\wo{i}$, namely, such that $j<i$ and $i+\card{\wo{i}}\le
j+\card{\wo{j}}$. For any $t \geq 1$, the following invariants holds
(where $\act{0}$ is trivially the set of all the suffixes):
\begin{enumroman}
\item\label{inv:presel} [\textsf{prefixes}]: $\presel{t-1}$ and
  $\presel{t}$ are prefixes of the (unknown) $k$th smallest suffix of
  $S$, and $\card{\presel{t-1}} < \card{\presel{t}}$.

\item\label{inv:work:ext} [\textsf{\wk{}s}]: For any suffix $\suf{i}$,
  its \wk{} $\wo{i}$ is either degenerate (a single mismatching
  symbol) or $\wo{i} = \presel{t'}$ for a phase $t' \leq
  t$. Moreover, $\wo{i}=\presel{t}$ iff $\suf{i}\in\act{t}$.

\item\label{inv:compare} [\textsf{comparing}]: For any $\suf{i}$ and
  $\suf{j}$, $\card{\wo{i}} \ne\card{\wo{j}}$ implies that we know
  whether $\wo{i} < \wo{j}$ or $\wo{i} > \wo{j}$.

\item\label{inv:nesting} [\textsf{nesting}]: For any two
  suffixes $\suf{i}$ and $\suf{j}$, their \wk{}s $\wo{i}$ and
  $\wo{j}$ do not overlap (either they are disjoint or one is
  contained within the other). Namely, $i>j$ implies
  $i+\card{\wo{i}}\le j+\card{\wo{j}}$ or $i \ge
  j+\card{\wo{j}}$.

\item\label{inv:covering} [\textsf{covering}]: The \wk{}s of the active
  suffixes are all maximal and, together with the maximal \wk{}s
  generated by the inactive suffixes, form a \emph{non-overlapping
    covering} of $S$ (i.e.\mbox{} $S =
  \wo{i_1}\wo{i_2}\cdots\wo{i_r}$, where $i_1 < i_2 < \cdots <
  i_r$ and either $\suf{i_j} \in \act{t}$, or $\suf{i_j} \in
  \inact{t}$ and $\wo{i_j}$ is maximal, for $1 \leq j \leq r$).
\end{enumroman}

\begin{lemma}\label{lem:alg:inv}
  After phase transition $(t-1 \rightarrow t)$ with $t \ge 1$, the
  invariants~\ref{inv:presel}--\ref{inv:covering} are maintained.
\end{lemma}


\begin{theorem}\label{theo:alg:correct}
  The algorithm terminates in a phase $t \leq N$, and returns the $k$th
  lexicographically smallest suffix.
\end{theorem}

\begin{theorem}\label{theo:compl}
  Our suffix selection algorithm requires $\Oh{N}$ time in the worst
  case.
\end{theorem}

This simpler suffix selection
algorithm is still cache ``unfriendly''. For example, 
it requires $\Oh{N}$ block transfers with a
string $S$ with period
length $\Thetah{B}$ (if $S$ is a prefix of $g^i$ for some integer $i$, then
$g$ is a period of $S$).

\vspace*{-2ex}
\section{Cache-Aware Sparse Suffix Selection}
\label{sec:sparse:selection}

In \emph{the sparse suffix selection problem}, along with the string
$S$ and the rank $k$ of the suffix to retrieve, we are also given a
set $\reduced$ of suffixes such that $\card{\reduced}=\Oh{N/B}$ and
the $k$th smallest suffix belongs in $\reduced$. We want to find the
wanted suffix in $\Oh{N/B}$ block transfers using the ideas of the
algorithm described in Section~\ref{sec:simple}.

Consider first a particular situation in which the suffixes are
equally spaced $B$ positions each other. We can split $S$ into blocks
of size $B$, so that $S$ is conceptually a string of $N/B$
metacharacters and each suffix starts with a metacharacters. This is a
fortunate situation since we can apply the algorithm described
in Section~\ref{sec:simple} as is, and solve the problem in the
claimed bound. The nontrivial case is when the suffixes can be in arbitrary
positions.

Hence, we revisit the algorithm described in
Section~\ref{sec:simple} to make it more cache efficient. Instead
of trying to extend the \wk{} of an active suffix $\suf{i}$ by just
using the \wk{}s of the following inactive suffixes, we  try to
batch these \wk{}s in a sufficiently long segment, called
\emph{reach}. Intuitively, in a step similar to step~\ref{step:select}
of the algorithm in Section~\ref{sec:simple}, we could first apply the
\SEL\ algorithm to the set of reaches. Then, after we select a subset
of equal reaches, and the corresponding subset of active suffixes, we
could extend their \wk{}s using their reaches. This could cause
collisions between the suffixes and they could be managed in a way
similar to what we did in Section~\ref{sec:simple}. This yields
the notion of super-phase transition.

\bsubsection{Super-phase transition} 
\label{sub:super-phase-transition}

The purpose of a super-phase is to group consecutive phases together,
so that we maintain the same invariants as those defined in
Section~\ref{subsec:inv}.  However, we need further concepts to
describe the transition between super-phases.  We number the
super-phases according to the numbering of phases. We call a
super-phase $m$ if the \emph{first} phase in it is $m$ (in the overall
numbering of phases).

\iparagraph{Reaches, pseudo-collisions and prospective reaches.} 
Consider a generic super-phase $m$. Recall that, by the
invariant~\ref{inv:covering} in Section~\ref{subsec:inv}, the phase
transitions maintain the string $S$ partitioned into maximal \wk{}s.
We need to define a way to access enough (but not too many)
consecutive ``lookahead'' \wk{}s following each active suffix, before running
the
super-phase. Since some of these active suffixes will become inactive
during the phases that form the super-phase, we cannot prefetch too
many such \wk{}s (and we cannot predict which ones will be
effectively needed). This idea of prefetching leads to the following
notion.

For any active suffix $\suf{i}\in\act{m}$, the {\em reach} of
$\suf{i}$, denoted by $\reach{i}$, is the
\emph{maximal} sequence of consecutive \wk{}s
$\wo{l_1}\wo{l_2}\cdots\wo{l_f}$ such that
\begin{enumroman}
\item $i < l_1 < l_2 < \cdots < l_f$ and $l_f-l_1 < B$;
\item $\wo{i}$ and $\wo{l_1}$ are adjacent and, for  $1<x\le f$,
  $\wo{l_{x-1}}$ and $\wo{l_{x}}$ are adjacent in $S$;
\item \label{item:nofull} if $\suf{j}$ is the leftmost active suffix
  in $S[i+1\ldots N]$, then $l_f \leq j$.
\end{enumroman}

We call a reach \emph{full} if $l_f < j$ in
condition~\ref{item:nofull}, namely, we do not meet an active suffix
while loading the reach. Since we know how to compare two \wk{}s, we
also know how to compare any two reaches $\reach{i},\reach{j}$, seen
as sequences of \wk{}s. We have the following.

\begin{lemma}\label{lem:reach:noprefix}
  For any two reaches $\reach{i}$ and $\reach{j}$, such that
  $\card{\reach{i}}<\card{\reach{j}}$, we have that $\reach{i}$ cannot
  be a prefix of $\reach{j}$.
\end{lemma}

Using reaches, we must possibly handle the collisions that may occur
in an arbitrary phase that is internal to the current super-phase. We
therefore introduce a notion of collision for reaches that is called
pseudo-collision because it does not necessarily implies a collision.

For any two reaches $\reach{i},\reach{j}$ such that $i<j$, we say that
$\reach{i}$ and $\reach{j}$ \emph{pseudo-collide} if
$\reach{i}=\reach{j}$ and the last \wk{} of $\reach{i}$ \emph{is}
$\wo{j}$ itself (not just \emph{equal to} $\wo{j}$).  Thus, the
last \wk{} of $\reach{j}$ is active and equal to $\wo{i}$ and
$\wo{j}$. Certainly, the fact that $\reach{i}$ and $\reach{j}$
pseudo-collide during a super-phase does not necessarily imply that
the \wk{}s $\wo{i}$ and $\wo{j}$ collide in one of its phases.  A
{\em pseudo-collision} $\pseudocollis{l}$ is a maximal sequence
$\reach{l_1}\reach{l_2}\cdots\reach{l_a}$ such that $\reach{l_f}$ and
$\reach{l_{f+1}}$ pseudo-collide, for any $1\le f<a$.  For our
algorithm, a degenerate pseudo-collision is a sequence of just
one reach.

Let us consider an active suffix $\suf{i}$ and the pseudo-collision to
which $\reach{i}$ belongs. Let us suppose that the pseudo-collision is
$\reach{l_1}\reach{l_2}\cdots\reach{l_{f-1}}\reach{i}\reach{l_{f+1}}
\cdots\reach{ l_a }$ (i.e.\mbox{} $\reach{i}$ is the $f$th reach).
Also, let us consider the reach $\reach{u}$ of the last \wk{}
$\wo{u}$ that appears in $\reach{l_a}$ (by the definition of
pseudo-collision, we know that the last \wk{} $\wo{u}$ of
$\reach{l_a}$ is equal to its first \wk{}, so $\suf{u}$ is active and
has a reach).  The {\em prospective reach} of an active \wk{}
$\wo{i}$, denoted by $\preach{i}$, is the sequence
$\reach{i}\reach{l_{f+1}}\cdots\reach{l_a}\tailpr{i}$, where
$\tailpr{i}=\lcp{\reach{i},\reach{u}}$ is the \emph{tail} of
$\preach{i}$ and denotes the longest initial sequence of \wk{}s that is
common to both $\reach{i}$ and $\reach{u}$.  Analogously to
prospective \wk{}s, we can define a total order on the prospective
reaches.  The \emph{multiplicity} of $\preach{i}$, denoted by
$\multpreach{i}$, is $a-f+1$ (that is the number of reaches following
$\reach{i}$ in the pseudo-collision plus $\reach{i}$). 

\begin{lemma}\label{lem:preach:comp}
  If the invariants for the phases hold for the current super-phase
  then, for any two reaches $\reach{i}$ and $\reach{j}$ such that
  $\reach{i}=\reach{j}$, we have that their prospective reaches
  $\preach{i}$ and $\preach{j}$ can be compared in $\Oh{1}$ time,
  provided we know the lengths of $\tailpr{i}$ and $\tailpr{j}$.
\end{lemma}

\iparagraph{Super-phase transition  $(m \rightarrow m')$.}
The transition from a super-phase $m$ to the next super-phase~$m'$
\emph{emulates} what happens with phases $m,m+1,\ldots,m'$ in the
algorithm of Section~\ref{sec:simple}, but using $O(N/B)$ block
transfers.

\begin{enumerate}

\item\label{step:super:findreach} For each active suffix $\suf{i}$, we
  create a pointer to its reach $\reach{i}$.

\item
  \label{step:super:firstselect}
  We find the $(k-\prerank{m})$th lexicographically smallest reach
  $\rho$ using \SEL\ on the $\Oh{N/B}$ pointers to reaches created in
  the previous step. The sets
  $\mathscr{R}_{=}=\{\suf{i}\;\vert\;\suf{i}$ is active and
  $\reach{i}=\rho\}$,
  $\mathscr{R}_{<}=\{\suf{i}\;\vert\;\suf{i}$ is active and
    $\reach{i}<\rho\}$, and
  $\mathscr{R}_{>}=\{\suf{i}\;\vert\;\suf{i}$ is active and
    $\reach{i}>\rho\}$ are thus identified, and, for any
  $\suf{i}\in\mathscr{R}_{<}\cup\mathscr{R}_{>}$, the length of
  $\lcp{\reach{i},\rho}$.\footnote{Given strings $S$ and $T$, their longest common prefix $\lcp{S,T}$ 
is longest string $U$ such that both $S$ and $T$ start with $U$.}  If $\card{\mathscr{R}_=}=1$, 
we stop and
  return $\suf{i}$, such that $\suf{i}\in\mathscr{R}_=$, as the $k$th
  smallest suffix in~$S$.

\item
  \label{step:super:preach:build} 
  For any $\suf{i}\in\mathscr{R}_{=}$, we compute its prospective
  reach $\preach{i}$.

\item
  \label{step:super:secondselect}
  We find the $(k-\prerank{m}-\card{\mathscr{R}_{<}})$th
  lexicographically smallest prospective reach $\pi$ among the ones in
  $\set{\preach{i}\;\vert\;\suf{i}\in\mathscr{R}_{=}}$, thus obtaining
  $\mathscr{P}_{=}=\set{\suf{i}\;\vert\;\suf{i}\mbox{ is active and
    }\preach{i}=\pi}$,
  $\mathscr{P}_{<}=\set{\suf{i}\;\vert\;\suf{i}\mbox{ is active and
    }\preach{i}<\pi}$,
  $\mathscr{P}_{>}=\set{\suf{i}\;\vert\;\suf{i}\mbox{ is active and
    }\preach{i}>\pi}$, and, for any
  $\suf{i}\in\mathscr{P}_{<}\cup\mathscr{P}_{>}$, the length of
  $\lcp{\preach{i},\pi}$.  If $\card{\mathscr{P}_=}=1$, we stop and
  return $\suf{i}$, such that $\suf{i}\in\mathscr{P}_=$, as the $k$th
  smallest suffix in~$S$.
\end{enumerate}

\begin{theorem}
  \label{theorem:ext-mem-sparse-suffix}
  The sparse suffix selection problem can be solved using $\Oh{N/B}$
  block transfers in the worst case.
\end{theorem}

\vspace*{-3ex}
\section{Optimal Cache-Aware Suffix Selection}
\label{sec:overview-external-memory}
\label{sec:redux}

The approach in Sec.~\ref{sec:sparse:selection} does not work 
if the number of input active suffixes is $\omegah{N/B}$. The
process would cost $\Oh{\frac{N}{B}\log B}$
block transfers (since it would take $\Omegah{\log B}$ transitions to
finally have $\Oh{N/B}$ active suffixes left). 
However, if we were able to find a set $\reduced$ of $\Oh{N/B}$ suffixes
such that one of them is the $k$th smallest, we could solve the
problem with $\Oh{N/B}$ block transfers using the algorithm in
Sec.~\ref{sec:sparse:selection}. In this section we show how to compute such
a set $\reduced$.

Basically, we consider all the substrings of length $B$ of $S$ and we
select a suitable set
of $\pivot > B$ pivot substrings that are roughly evenly spaced. Then, we find
the pivot that is lexicographically
``closest'' to the wanted $k$-th and one of the following two situations arises:

$\bullet$~~We are able to infer that the $k$th smallest suffix is strictly between
two
  consecutive pivots (that is its corresponding substring of $B$ characters is
strictly greater and smaller of the two pivots). In this case, we return
  all the $\Oh{N/\pivot}=\Oh{N/B}$ suffixes that are contained between the two
  pivots. 


$\bullet$~~We can identify the suffixes that have the first $B$ characters
  equal to those of the $k$th smallest suffix. We show that, in case they
  are still $\Omegah{N/B}$ in number, they must satisfy some periodicity
  property, so that we can reduce them to just $\Oh{N/B}$ with
  additional $\Oh{N/B}$ block transfers.


\vspace*{-2ex}
\subsection{Finding pivots and the key suffixes}\label{subsec:pivots}

\noindent{}Let $\pivot=\sqrt{\frac{M^c}{B}}$, for a suitable constant $c>1$. We
proceed with the following steps.

\emph{First.} We sort the first $M^c$ substrings of length $B$ of $S$ (that is 
substrings $S[1\ldots B]$, $S[2\ldots B+1]$,\dots, $S[M^c-1\ldots
B+M^c-2]$, $S[M^c\ldots B+M^c-1]$). Then we sort the second $M^c$ substrings of
length $B$ and so forth until all the $N$ positions in $S$ have been considered.
The product of this step is an array $V$ of $N$ pointers to the
substrings of length $B$ of $S$.

\emph{Second.} We scan $V$ and we collect in an array $U$ of $N/\pivot$ positions the
$N/\pivot$ pointers  $V[\pivot],V[2\pivot],V[3\pivot],\ldots$.

\emph{Third.} We (multi)-select from $U$ the $\pivot$ pointers to the substrings (of
length $B$) $b_1,\ldots,b_\pivot$ such that $b_i$ has rank
$i\frac{N}{\pivot^2}$ among the substrings (pointed by the pointers) in $U$.  
These are the pivots we were looking for. We store the $\pivot$ (pointers to
the) pivots in an array $U'$.

\emph{Fourth.} We need to find the rightmost pivot $b_x$ such that the number of
substrings (of length $B$ of $S$) lexicographically smaller than $b_x$ is less
than $k$ (the rank of the wanted suffix). 
We cannot simply distribute all the substrings of length $B$ according to all
the $\pivot$ pivots in $U'$, because it would be too costly.   
Instead, we proceed with the following refining strategy.  

\begin{itemize}
\item[$1$.] From the $\pivot$ pivots in $U'$ we extract the group $G_1$ of
$\delta M$ equidistant pivots, where $\delta<1$ is a suitable constant, (i.e.
the pivots $b_{t},b_{2t},\ldots$, where $t=\frac{\pivot}{\delta M}$).
Then, for any $b_{j}\in G_{1}$, we find out how many substrings of size $B$ are
lexicographically smaller than $b_j$. After that we find the  rightmost pivot
$b_{x_1}\in G_1$ such that the number of
substrings (of length $B$) smaller than $b_{x_1}$ is less
than $k$. 
\item[$2$.] From the $\frac{\pivot}{\delta M}$ pivots in $U'$ following
$b_{x_1}$ we extract the group $G_2$ of $\delta M$ equidistant pivots.  Then,
for any $b_{j}\in G_{2}$, we find out how many substrings of size $B$ are
smaller than $b_j$. After that we find the  rightmost pivot
$b_{x_2}\in G_2$ such that the number of
substrings smaller than $b_{x_2}$ is less
than $k$. 
\end{itemize}
\qquad More generally:
\begin{itemize}
\item[$f$.] Let $G_{f}$ be the $\delta M$ pivots in $U'$ following
$b_{x_{f-1}}$. Then,
for any $b_{j}\in G_{f}$, we find out how many substrings of size $B$ are
smaller than $b_j$. After that we find the  rightmost pivot
$b_{x_{f}}\in G_f$ such that the number of
substrings smaller than $b_{x_{f}}$ is less
than $k$. 
\end{itemize}

The pivot $b_{x_f}$ found in the last iteration is the pivot $b_x$
we are looking for in this step.

\emph{Fifth.} We scan $S$ and compute the following two numbers: the number
$n^{\scriptscriptstyle<}_{x}$ of substrings of length $B$
lexicographically smaller than $b_x$; the number
$n^{\scriptscriptstyle=}_{x}$ of substrings equal to $b_x$.

\emph{Sixth.} In this step we treat the following 
case: $n^{\scriptscriptstyle<}_{x}< k\le n^{\scriptscriptstyle<}_{x}+
n^{\scriptscriptstyle=}_{x}$. 
More specifically, this implies that the wanted
$k$th smallest suffix has its prefix of $B$ characters equal to $b_x$.
We proceed as follows. We scan $S$ and gather in a contiguous zone $R$
(the
indexes of) the suffixes of
$S$ having their prefixes of $B$ characters equal to $b_x$.
In this case we
have already found the key suffixes (whose indexes reside in $R$). Therefore the
computation
in this section ends here and we proceed to discard some of
them (sec.~\ref{subsec:discard}).
 
\emph{Seventh.} In this step we treat the following remaining 
case: $n^{\scriptscriptstyle<}_{x}+n^{\scriptscriptstyle=}_{x}< k$. 
In other words, in this case we know that the prefix 
of $B$ characters of the wanted $k$th smallest suffix is
(lexicographically) greater than $b_{x}$ 
and smaller than $b_{x+1}$. 
Therefore, we scan $S$ and gather in a contiguous zone $R$ (the
indexes of) the suffixes of
$S$ having their prefix of $B$ characters greater than $b_{x}$ 
and smaller than $b_{x+1}$.
Since there are less than $N/B$ such suffixes (see below
Lemma~\ref{lem:pivoting}), we have
already
found the set of sparse active suffixes (whose indexes reside in $R$)
that will be processed in Sec.~\ref{sec:sparse:selection}.

\begin{lemma}\label{lem:pivoting}
For any $S$ and $k$, either the number of key suffixes found is $\Oh{N/B}$, or
their prefixes of $B$ characters are all the same. 
\end{lemma}

\begin{lemma}\label{lem:pivoting:compl}
Under the tall-cache assumption, finding the key suffixes needs
$\Oh{N/B}$ block transfers in the worst case.
\end{lemma}

\vspace*{-3ex}
\subsection{Discarding key suffixes}\label{subsec:discard}
Finally, let us show how to reduce the number of key suffixes
gathered in Sec.~\ref{subsec:pivots} to $\le 2N/B$ so that we can pass them
to the sparse suffix selection algorithm (Sec.~\ref{sec:sparse:selection}). 
Let us assume that the number of key suffixes is greater than $2N/B$.

The indexes of the key suffixes have been previously stored in an array $R$.
Clearly, the $k$th smallest suffix is among the ones in $R$. We also
know the number $n^{\scriptscriptstyle<}$ of suffixes of $S$ that
are lexicographically smaller than each suffix in $R$.
Finally, we know that there exists a string $\fblock$ of length $B$ such that
$R$ \emph{contains all and only the suffixes $\suf{i}$ such that the prefix of
length $B$ of $\suf{i}$ is equal to $\fblock$} (i.e. $R$ contains the indexes
of all the occurrences of $\fblock$ in $S$). 

To achieve our goal we exploit the possible periodicity of the string $\fblock$.
A string $u$ is \emph{a period} of a string $v$ ($\card{u}\le\card{v}$) if $v$
is a prefix of $u^i$ for some integer $i\ge 1$. \emph{The period} of $v$ is the
smallest of its periods. We exploit the following:

\begin{property}[\cite{galil}]\label{prop:period}
If $\fblock$ occurs in two positions $i$ and $j$ of $S$ and
$0<j-i<\card{\fblock}$ then $\fblock$ has a period of length $j-i$.
\end{property}
 
\noindent{}Let $\period$ be the period of $\fblock$. Since the number of
suffixes in $R$ is greater than $2N/B$, \emph{there must be some overlapping}
between the occurrences of $\fblock$ in $S$. Therefore, by
Property~\ref{prop:period}, we can conclude that
$\card{\period}<\card{\fblock}$.
For the sake of presentation let us assume that $\card{\fblock}$ \emph{is not a
multiple of} $\card{\period}$ (the other case is analogous).

From how $R$ has been built (by left to right scanning of $S$) we know that
the indexes in it are in increasing order, that is $R[i]<R[i+1]$,
for any $i$ (i.e. the indexes in $R$ follow the order, from left to right, in
which the corresponding suffixes may be found in $S$). 
Let us consider a
\emph{maximal subsequence} $R_i$ of $R$ such that, for any $1\le j<\card{R_i}$,
$R_i[j+1]-R_i[j]\le B/2$ (i.e. the occurrence of 
$\fblock$ in $S$ starting in position $R_i[j]$ overlaps the one 
starting in position $R_i[j+1]$ by at least $B/2$ positions). Clearly, any two
of these subsequences of $R$ do
not overlap and hence $R$ can be
seen as the concatenation $R_1R_2\cdots$ of these subsequences.
From the definition of the partitioning of $R$ and from the periodicity
of $\fblock$ we have:

\begin{lemma}\label{lem:key:collis}
The following statements hold:
\begin{enumroman}
\item There are less than $2N/B$ such subsequences.
\item For any $R_i$, the substring $S[R_i[1]\ldots R_i[\card{R_i}]+B-1]$ (the
substring of $S$ spanned by the substrings whose indexes are in $R_i$) has
period $\period$.
\item The substring of length $B$ of $S$ starting in position
$R_i[\card{R_i}]+\card{\period}$ (the substring starting one period-length
past the rightmost member of $R_i$) is not equal to $\fblock$. 
\end{enumroman}
\end{lemma}

For any key suffix $\suf{j}$, let us consider the following prefix:
$\prefi{j}=S[j\ldots R_i[\card{R_i}]+\card{\period}+B-1]$, where $R_i$ is the
subsequence of $R$ where (the index of) $\suf{j}$ belongs to.
By Lemma~\ref{lem:key:collis}, we know two things about $\prefi{j}$:
$(a)$ the prefix  of
length $\card{\prefi{j}}-\card{\period}$ of $\prefi{j}$ has period $\period$;
$(b)$ the suffix of length $B$ of $\prefi{j}$ \emph{is not equal} to $\fblock$.

In light of this, we associate with any
key suffix $\suf{j}$ a pair of integers $\seq{\firstint{j},\secondint{j}}$
defined as follows: $\firstint{j}$ is equal to \emph{the number of
complete periods} $\period$ in the prefix of
length $\card{\prefi{j}}-\card{\period}$ of $\prefi{j}$; $\secondint{j}$ is
equal to $\card{R_i}+\card{\period}$ (that is the index of the substring of
length $B$ starting one period-length past the rightmost member of $R_i$).

There is natural total order $\ordrelation$ that can be defined over the key
suffixes. It is based on the pairs of integers
$\seq{\firstint{j},\secondint{j}}$ and it is defined as follow. 
For any two key suffixes $\suf{j'},\suf{j''}$:
\begin{itemize}
\item If $\firstint{j'}=\firstint{j''}$ then  $\suf{j'}$ and $\suf{j''}$ are
equal (according to $\ordrelation$).
\item If $\firstint{j'}<\firstint{j''}$ then $\suf{j'}\ordrelation\suf{j''}$ iff
$S[\secondint{j'}\ldots \secondint{j'}+B-1]$ is lexicographically smaller than~$\fblock$. 
\end{itemize}
By Lemma~\ref{lem:key:collis}, we know that the  suffix of length $B$ of
$\prefi{j'}$ (which is the substring $S[\secondint{j'}\ldots
\secondint{j'}+B-1]$) \emph{is not equal} to $\fblock$. Therefore the total
order $\ordrelation$ is well defined.

We are now ready to describe the process for reducing the number of key
suffixes. We proceed with the following steps.

\emph{First.} By scanning $S$ and $R$, we compute the pair
$\seq{\firstint{j},\secondint{j}}$ for any key suffix $\suf{j}$. The pairs are
stored in an array (of pairs of integers) $\pairs$.

\emph{Second.} We scan $S$ and compute the array $\compar$ of $N$ positions
defined as follows: for any $1\le i\le N$, $\compar[i]$ is equal to $-1$, $0$ or
$1$ if $S[i\ldots i+B-1]$ is less than, equal to or greater than $\fblock$,
respectively (the array $\compar$ tells us what is the result of the
comparison of $\fblock$ with any substring of size $B$ different from it).

\emph{Third.} By scanning $\pairs$ and $\compar$ at the same time, we compute the
array $\pairscomp$ of size $\card{\pairs}$, such that, 
for any $l$, $\pairscomp[l]=\compar[\pairs[l].\secondint{}]]$ (where
$\pairs[l].\secondint{}$ is the second member of the pair of integers in
position $l$ of $\pairs$).

\emph{Fourth.} Using $\pairs$ and $\pairscomp$, we select the
$(k-n^{\scriptscriptstyle<})$-th smallest key suffix $\suf{x}$ and all the
key suffixes equal to $\suf{x}$ \emph{according to the total order}
$\ordrelation$ (where $n^{\scriptscriptstyle<}$ is the number of suffixes of $S$
that are lexicographically smaller than each suffix in $R$, known since
Sec.~\ref{subsec:pivots}). The set of the selected key suffixes is the output
of the process.

\begin{lemma}\label{lem:discar:howmany}
 At the end of the discarding process, the selected key suffixes are less than
$2N/B$ in number and the $k$th lexicographically smallest suffix is among them. 
\end{lemma}

\begin{lemma}\label{lem:discar:compl}
The discarding process requires $\Oh{N/B}$ block transfers at the worst case. 
\end{lemma}

\begin{theorem}
  \label{theorem:ext-mem-general-alphabet}
  The suffix selection problem for a string defined over a general
  alphabet can be solved using $\Oh{N/B}$ block transfers in the worst
  case.
\end{theorem}




\vspace*{-3ex}
\newcommand{\lbk}{\linebreak[0]}\def\softl{l\kern -0.45ex\raise
  0.1ex\hbox{'}\kern -0.10ex}%

\end{document}